\documentclass[a4paper]{jpconf}
\usepackage{graphicx}
\usepackage{amsmath}
\usepackage{amssymb}
\usepackage{mathrsfs}
\usepackage{extarrows}
\usepackage{fancyhdr}
\fancypagestyle{firstpage}{
  \lhead{Journal of Physics: Conference Series, accepted (2021)
	\newline
	The proceedings of the International conference PhysicA.SPb/2021}
  \rhead{}
}

\begin{document}
\title{The candidates for Class I methanol masers}

\author{A V Nesterenok}

\address{Ioffe Institute, 26 Polytechnicheskaya St., 194021, Saint Petersburg, Russia}

\ead{alex-n10@yandex.ru}
\thispagestyle{firstpage}  

\begin{abstract}
The collisional excitation of methanol molecule in non-dissociative magnetohydrodynamic shock waves is considered. All essential chemical processes that determine methanol abundance in the gas are taken into account in the shock model. The large velocity gradient approximation is used in the calculations of energy level populations of the molecule. We calculate the optical depth for inverted methanol transitions, and present the list of candidates for Class I methanol masers that have collisional pumping mechanism.
\end{abstract}

\section{Introduction}
Maser emission from methanol was first discovered by \cite{Barrett1971} towards Orion-KL region in the transitions of the series E~$J_2 \to J_1$ near 25~GHz ($J = 4,5,6,7,8$). Orion remained the only known methanol maser source until mid-1980's -- new methanol maser transition A$^+$~$9_2 \to 10_1$ at frequency 23.121~GHz was discovered towards compact H~II regions W3(OH) and possibly NGC~7538 \cite{Wilson1984}. Further, \cite{Wilson1985} reported on the detection of E~$2_1 \to 3_0$ methanol transition at frequency 19.967~GHz towards W3(OH). Maser transitions E~$4_{-1} \to 3_0$ at 36.169~GHz and A$^+$~$7_0 \to 6_1$ at 44.069~GHz were discovered in Sgr~B2, the second transition was detected also in W51 and other two galactic sources \cite{Morimoto1985}. The 100-m telescope was used by \cite{Menten1986} to observe the transitions of the series E~$J_2 \to J_1$ near 25~GHz towards a selection of galactic molecular-line sources. They found that sources of maser emission in these transitions often appear separated from known radio or infrared sources. 

Strong methanol masers in the E~$2_0 \to 3_{-1}$ transition at 12.178~GHz were discovered in a number of galactic sources including W3(OH) \cite{Batrla1987}. Based on the observational data on methanol masers available at that time, \cite{Batrla1987} suggested that methanol masers can be divided into two categories. Sources that show maser emission in the $J_2 \to J_1$ series of E-type methanol belong to the first category or class of masers. These masers can be found apart from OH and H$_2$O masers, compact H~II regions and infrared sources. Second category of masers are found close to compact H~II regions. The known at that time methanol transitions at 12.178, 19.967 and 23.121~GHz belong to this category of masers. Surveys of 36.169 and 44.069~GHz methanol transitions towards galactic star-forming regions \cite{Haschick1989,Bachiller1990,Haschick1990}, and interferometric observations of methanol transitions E~$5_{-1} \to 4_0$ at 84.521~GHz and A$^+$~$8_0 \to 7_1$ at 95.169~GHz \cite{Batrla1988,Plambeck1990} revealed that all these lines belong to the first category of masers. The two categories of methanol masers were subsequently labelled as Class I and Class II \cite{Menten1991b}. 

The separation of methanol transitions into two classes corresponds to different maser pumping mechanisms but not to the type of associated astronomical objects -- Class I methanol masers have the collisional pumping mechanism, while Class II methanol masers are pumped by radiative excitation \cite{Cragg1992,Slysh1994}. The Class I methanol masers trace shocked gas in star-forming regions, and the shocks may be attributed to different phenomena of star-formation as bipolar outflows or expanding H II regions \cite{Voronkov2014}. The interstellar shocks compress and heat up the gas, which is a conducive condition for the collisional pumping of Class I methanol masers. Here, we use the magnetohydrodynamic model of C-type shock wave published by \cite{Nesterenok2018,Nesterenok2019} to study the collisional excitation of methanol and emergence of maser radiation. In this work we provide the list of methanol transitions that have population inversion in the post-shock region and may be observed as masers.

\section{The C-type shock model}
The C-type shock ('C' from the word 'continuous') is formed when the speed of the gas flow is lower than the magnetosonic speed, but higher than the speed of sound for the neutral gas component. In this case, the medium upstream from the shock front is disturbed by magnetosonic waves that propagate faster than the gas flow speed. There is a compression of the ionic gas component and the magnetic field upstream from the shock front. An abrupt, nearly discontinuous shock front becomes smooth due to the ambipolar diffusion of ions and neutral gas component. For gas densities and ionization fractions expected in dark molecular clouds, the maximal speed of C-type shock is about 40--60~km~s$^{-1}$. At higher shock speeds, the molecular hydrogen (the main coolant in hot/warm molecular gas) is dissociated and gas-dynamic shock discontinuity forms. A detailed description of the model of a steady-state C-type shock used in this work is provided by \cite{Nesterenok2018,Nesterenok2019,Nesterenok2021}. All important chemical reactions that affect the methanol abundance in the shock wave are taken into account.

The shock wave simulations consist of two steps: (1) the simulations of the chemical evolution of a dark molecular cloud and (2) the calculations of the shock propagation. The key physical parameters of the shock model are given in the table~\ref{table_parameters}.

\begin{table}
\caption{\label{table_parameters}Parameters of the shock model.}
\begin{center}
\begin{tabular}{p{7.5cm} p{3.5cm}}
\hline
Parameter & Value \\
\hline 
Pre-shock gas density, $n_{\mathrm{H_2}}$ & $10^4$, $10^5$, $10^6$~cm$^{-3}$\\
Shock speed, $u_{\text{s}}$ & $17.5$~km~s$^{-1}$ \\
Cosmic ray ionization rate, $\zeta_\mathrm{H_2}$ & $3 \times 10^{-17}$~s$^{-1}$ \\
Visual extinction, $A_V$ & 10 \\
Micro-turbulence speed, $v_{\text{turb}}$ & 0.3~km~s$^{-1}$ \\
Methanol abundance at the shock start with respect to H nucleus number density & $10^{-5}$ \\
\hline
\end{tabular}
\end{center}
\end{table}

\section{The maser modelling}
The profiles of physical parameters and abundances of chemical species which are derived from shock wave simulations are used in calculations of energy level populations of methanol molecule. The shock profile is split into layers, and energy level populations are calculated for each layer. The system of equations for the energy level populations $n_{\text{i}}$ is

{\setlength{\mathindent}{0pt}
\begin{equation}
\begin{array}{c}
\displaystyle
\sum_{k=1, \, k \ne i}^N \left( R_{\mathrm{ki}} + C_{\mathrm{ki}} \right) n_{\mathrm{k}} - n_{\mathrm{i}} \sum_{k=1, \, k \ne i}^N \left( R_{\mathrm{ik}} + C_{\mathrm{ik}} \right)=0, \quad i=1,...,N-1, \\ [15pt]
\displaystyle
\sum_{i=1}^N n_{\mathrm{i}} = 1,
\end{array}
\label{eq_stat}
\end{equation}}

\noindent
where $N$ is the total number of energy levels, $C_\text{ik}$ is the rate coefficient of collisional transition from level $i$ to level $k$, $R_\text{ik}$ is the rate coefficient for the radiative transition. The rate coefficients for downward and upward radiative transitions are

\begin{equation}
\displaystyle
R_\mathrm{ik}^{\downarrow} = B_\mathrm{ik} J_\mathrm{ik} + A_\mathrm{ik}, \quad R_\mathrm{ki}^{\uparrow} = B_\mathrm{ki} J_\mathrm{ik},
\label{eq_radiat_trans}
\end{equation}

\noindent
where $A_\text{ik}$ and $B_\text{ik}$, $B_\text{ki}$ are the Einstein coefficients; $J_\text{ik}$ is the radiation intensity in spectral line averaged over the line profile and the direction. The large velocity gradient or Sobolev approximation is used in calculations of spectral line intensities \cite{Nesterenok2020b}. The system of equations (\ref{eq_stat}) is non-linear and the solution is obtained iteratively. The convergence criterion for the iterative series is the condition on the relative change in energy level populations at iterative step $< 10^{-5}$. The spectroscopic data on methanol molecule are taken from \cite{Mekhtiev1999}. The rate coefficients for the excitation of methanol in collisions of methanol with He and H$_2$ were calculated by \cite{Rabli2010,Rabli2010b,Rabli2011}. Here, the A and E symmetry species of methanol are assumed to be equally abundant. 

The expression for the gain of a molecular line $i \to k$ in the gas--dust cloud with plane-parallel geometry is

\begin{equation}
\displaystyle
\gamma_\mathrm{ik}(z,\mu, \nu)=\frac{\lambda^2 }{8 \pi} A_\mathrm{ik} n_\mathrm{mol} \left(n_\mathrm{i}-\frac{g_\mathrm{i}}{g_\mathrm{k}} n_\mathrm{k} \right) \phi(z,\mu, \nu) - \kappa_\mathrm{c},
\label{eq_gain}
\end{equation}

\noindent
where $\mu$ is the $\text{cos}$ of the angle $\theta$ between the gas flow direction and the line of sight, $n_\text{mol}$ is the molecule concentration (A- or E-symmetry species of methanol), $g_\text{i}$ and $g_\text{k}$ are statistical weights of energy levels, $\kappa_\text{c}$ is the absorption coefficient of the dust. The spectral profile of the emission and absorption coefficients in the laboratory frame of reference is 

\begin{equation}
\phi(z,\mu, \nu) = \tilde{\phi}_\mathrm{ik} \left[ \nu - \nu_\mathrm{ik} \mu u(z) / c \right]
\label{eq_profile_lab}
\end{equation}

\noindent
where $\nu_\mathrm{ik}$ is the transition frequency, $u(z)$ is the gas velocity, $\tilde{\phi}_\mathrm{ik}(\nu)$ is the normalized spectral line profile in the co-moving frame of the gas. The optical depth in the given transition in the direction $\mu$ to the shock propagation is

\begin{equation}
\displaystyle
\vert \tau_{\mu}(\nu) \vert = \frac{1}{\mu}\int \, \mathrm{d}z \, \gamma_\mathrm{ik}(z,\mu, \nu),
\label{eq_tau}
\end{equation}

\noindent
where the integral is evaluated over the region with $\gamma_\mathrm{ik} > 0$. The expression (\ref{eq_tau}) attains maximum at some frequency -- the optical depth at the line centre. Here, we call the parameter $1/\mu$ as aspect ratio that is equal to the ratio of the amplification path of the maser ray to the shock width along the outflow direction. Usually, the value of aspect ratio of $\sim 10$ is taken in theoretical studies for strong masers. 

A detailed description of the calculations of energy level populations of methanol are given in our forthcoming paper \cite{Nesterenok2021}.

\section{Results}
Table~\ref{table_masers} shows the optical depth $\vert \tau_\text{1,max} \vert$ at the line centre in the direction of shock propagation ($\mu$ = 1) for a large sample of methanol transitions. All transitions that have optical depth $\vert \tau_\text{1,max} \vert \geq 0.1$ are shown in the table (the transition frequencies are truncated, not rounded). If we suggest that the shock wave is seen edge-on and aspect ratio is $1/\mu \approx 10$, the optical depth in maser transitions in question is $\vert \tau_\mu \vert > 1$. The calculations have been done for three pre-shock gas densities -- $n_\mathrm{H_2} = 10^4$, $10^5$, and $10^6$~cm$^{-3}$. In the post-shock region where maser emission originates, the gas density is few times higher due to gas compression and equals to about (2--10)$\times n_\mathrm{H_2}$. The maximum value of the optical depth achieved in these shock models is given in the table~\ref{table_masers}. The gas density at which the maximum of the optical depth is achieved is denoted by letters L, M and H for low, medium and high gas density, respectively. All known detected Class I methanol masers are included in the table~\ref{table_masers} -- the frequencies are shown in bold font for these transitions \cite{Voronkov2006,Towner2017,Ladeyschikov2019}. Methanol transitions are grouped into series, the most numerous transition series is E~$J_2 \to J_1$ near 25~GHz. Usually, the low-lying transitions are effectively pumped at low and medium gas densities. In particular, the widespread masers of A type methanol at 44.069, 95.169~GHz, and maser of E type methanol at 36.169~GHz are effectively pumped at low gas density. The higher the transition is located in the energy level diagram, the higher the gas density is necessary for the effective pumping of the maser, see table \ref{table_masers}. 

The detected masers near 25~GHz of E~$J_2 \to J_1$ series, 23.444~GHz of A$^{-}$~$J_1 \to (J-1)_2$ series, and two transitions at 9.9362 and 104.30~GHz of E~$J_{-1} \to (J-1)_{-2}$ series are rare masers. Previous theoretical models suggested that these transitions are inverted in high-density and high-temperature environment \cite{Leurini2018}. Indeed, the optimal condition for operation of these masers is the medium pre-shock gas density $n_\mathrm{H_2} = 10^5$~cm$^{-3}$, see table~\ref{table_masers}. According to our simulations, all transitions of the series E $J_2 \to (J-1)_1$ (including the detected lines at 218.44 and 266.83~GHz) are inverted in high-density gas despite the fact that these transitions have relatively low-energy upper levels (60~K for 266.83~GHz line). The optical depth in these transitions is low, $\vert \tau_\text{1,max} \vert \approx 0.1$, and the aspect ratio must be $1/\mu \approx 10$ to produce maser emission. Our results confirm the conclusions by \cite{Chen2019} who found that masers 218.44 and 266.83~GHz need high gas density for the pumping, $n_\mathrm{H_2} \sim 10^7$~cm$^{-3}$. The transitions of the series A~$J_1 \to J_2$ at 0.8342 and 2.5027~GHz (not detected yet as masers) are found to have appreciable optical depth at low gas density. 

\begin{table}
\caption{\label{table_masers}Predicted Class I methanol masers.}
\begin{center}
\begin{tabular}{llllllll}
\br
Transition & Frequency & Optical & Gas & Transition & Frequency & Optical & Gas\\
& GHz & depth & density & & GHz & depth & density\\
\mr
\multicolumn{4}{c}{A~$J_1 \to J_1$}          & \multicolumn{4}{c}{E~$J_2 \to (J-1)_3$}\\
A~$1_1 \to 1_1$ & 0.8342 & 0.2 & L             &E~$11_2 \to 10_3$ & 2.9259 & 0.4 & M\\
A~$2_1 \to 2_1$ & 2.5027 & 0.1 & L             &E~$12_2 \to 11_3$ & 51.759 & 0.2 & M\\
\multicolumn{4}{c}{A$^{+}$~$J_1 \to (J-1)_2$}  &E~$13_2 \to 12_3$ & 100.63 & 0.1 & H\\
A$^+$~$11_1 \to 10_2$ & 20.171 & 0.6 & M       &E~$14_2 \to 13_3$ & 149.53 & 0.1 & H\\
A$^+$~$12_1 \to 11_2$ & 62.906 & 0.4 & M       & \multicolumn{4}{c}{E~$J_{-1} \to (J-1)_{-2}$}\\
A$^+$~$13_1 \to 12_2$ & 105.06 & 0.2 & M       &E~$9_{-1} \to 8_{-2}$ & {\bf 9.9362} & 1.7 & M\\
\multicolumn{4}{c}{A$^{-}$~$J_1 \to (J-1)_2$}  &E~$10_{-1} \to 9_{-2}$ & 57.292 & 1.3 & M\\
A$^-$~$10_1 \to 9_2$ & {\bf 23.444} & 0.9 & M  &E~$11_{-1} \to 10_{-2}$ & {\bf 104.30} & 0.9 & M\\ 
A$^-$~$11_1 \to 10_2$ & 76.247 & 0.7 & M       &E~$12_{-1} \to 11_{-2}$ & 150.88 & 0.6 & M\\ 
A$^-$~$12_1 \to 11_2$ & 129.43 & 0.5 & M       &E~$13_{-1} \to 12_{-2}$ & 196.96 & 0.3 & M\\ 
A$^-$~$13_1 \to 12_2$ & 182.99 & 0.3 & M       &E~$14_{-1} \to 13_{-2}$ & 242.44 & 0.2 & M\\
A$^-$~$14_1 \to 13_2$ & 236.93 & 0.1 & H,M        & \multicolumn{4}{c}{E~$J_0 \to (J-1)_1$}\\ 
\multicolumn{4}{c}{A$^{+}$~$J_0 \to (J-1)_1$}  &E~$4_0 \to 3_1$ & 28.316 & 1.7 & L\\
A$^+$~$7_0 \to 6_1$ & {\bf 44.069} & 5.1 & L   &E~$5_0 \to 4_1$ & 76.509 & 1.1 & L\\
A$^+$~$8_0 \to 7_1$ & {\bf 95.169} & 2.1 & L,M &E~$6_0 \to 5_1$ & 124.56 & 0.6 & L,M\\ 
A$^+$~$9_0 \to 8_1$ & 146.61 & 1.4 & M          &E~$7_0 \to 6_1$ & 172.44 & 0.5 & M\\
A$^+$~$10_0 \to 9_1$ & 198.40 & 0.9 & M         &E~$8_0 \to 7_1$ & 220.07 & 0.3 & M\\ 
A$^+$~$11_0 \to 10_1$ & 250.50 & 0.4 & M        &E~$9_0 \to 8_1$ & 267.40 & 0.2 & M\\ 
A$^+$~$12_0 \to 11_1$ & 302.91 & 0.3 & H        &E~$10_0 \to 9_1$ & 314.35 & 0.1 & M\\ 
A$^+$~$13_0 \to 12_1$ & 355.60 & 0.2 & H        & \multicolumn{4}{c}{E~$J_{-1} \to (J-1)_0$}\\ 
\multicolumn{4}{c}{E~$J_2 \to J_1$}          &E~$4_{-1} \to 3_0$ & {\bf 36.169} & 3.3 & L \\
E~$3_2 \to 3_1$ & {\bf 24.928} & 1.1 & M     &E~$5_{-1} \to 4_0$ & {\bf 84.521} & 1.3 & M\\
E~$4_2 \to 4_1$ & {\bf 24.933} & 1.2 & M     &E~$6_{-1} \to 5_0$ & {\bf 132.89} & 0.8 & M\\
E~$2_2 \to 2_1$ & {\bf 24.934} & 0.6 & M     &E~$7_{-1} \to 6_0$ & 181.29 & 0.5 & M\\
E~$5_2 \to 5_1$ & {\bf 24.959} & 1.0 & M     &E~$8_{-1} \to 7_0$ & {\bf 229.75} & 0.3 & M,H\\
E~$6_2 \to 6_1$ & {\bf 25.018} & 0.8 & M     &E~$9_{-1} \to 8_0$ & {\bf 278.30} & 0.2 & H\\
E~$7_2 \to 7_1$ & {\bf 25.124} & 0.6 & M     &E~$10_{-1} \to 9_0$ & 326.96 & 0.2 & H\\
E~$8_2 \to 8_1$ & {\bf 25.294} & 0.4 & M     &E~$11_{-1} \to 10_0$ & 375.75 & 0.2 & H\\
E~$9_2 \to 9_1$ & {\bf 25.541} & 0.2 & M,H   &E~$12_{-1} \to 11_0$ & 424.72 & 0.1 & H\\
E~$10_2 \to 10_1$ & {\bf 25.878} & 0.2 & H   & \multicolumn{4}{c}{E~$J_2 \to (J-1)_1$}\\
E~$11_2 \to 11_1$ & 26.313 & 0.2 & H         &E~$2_2 \to 1_1$ & 121.68 & 0.2 & H \\
E~$12_2 \to 12_1$ & 26.847 & 0.2 & H         &E~$3_2 \to 2_1$ & 170.06 & 0.2 & H \\
E~$13_2 \to 13_1$ & 27.472 & 0.2 & H         &E~$4_2 \to 3_1$ & {\bf 218.44} & 0.1 & H \\
E~$14_2 \to 14_1$ & 28.169 & 0.2 & H         &E~$5_2 \to 4_1$ & {\bf 266.83} & 0.1 & H \\
\br
\end{tabular}
\end{center}
\end{table}

\begin{figure}[h]
\centering
\includegraphics[width = 9cm]{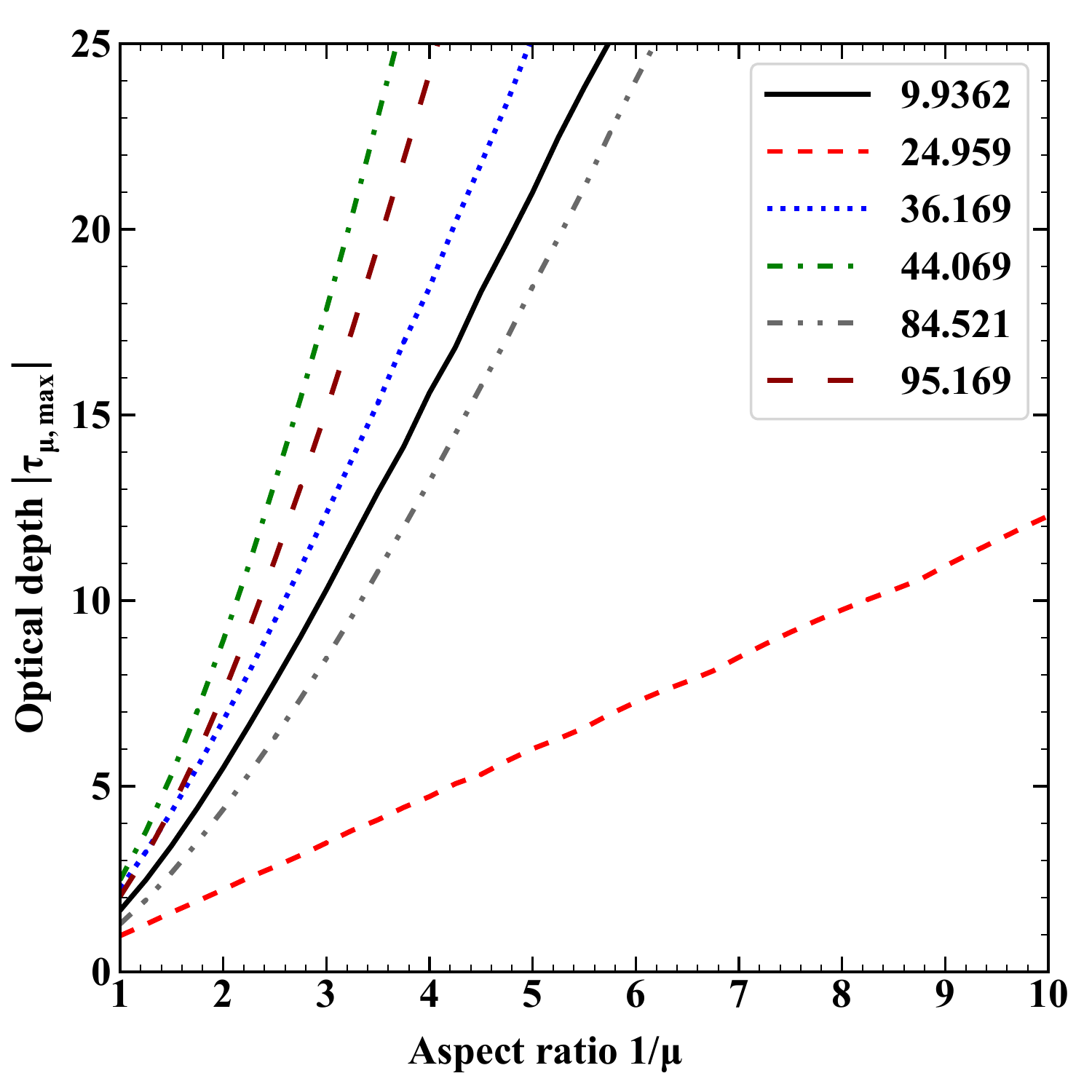}\hspace{1.5pc}
\begin{minipage}[b]{6cm}\caption{\label{fig_opt_depth}Optical depth at the line centre in methanol maser transitions as a function of the aspect ratio $1/\mu$, where $\mu$ is the \text{cos} of the angle between the line of sight and the direction of shock wave propagation. The results are shown for the shock model with pre-shock gas density $n_\mathrm{H_2} = 10^5$~cm$^{-3}$. The delegate of the transition series E~$J_2 \to J_1$ with $J$ = 5 at 24.959~GHz is shown.}
\end{minipage}
\end{figure}

Figure~\ref{fig_opt_depth} shows the optical depth at the line centre as a function of the aspect ratio $1/\mu$. The optical depth increases with decrease of $\mu$ faster than $\vert \tau_1 \vert/\mu$ as the frequency shift of the spectral profile across the shock region decreases with $\mu$ decrease, see equation (\ref{eq_profile_lab}). The radiation originated in different sides of the shock region becomes in resonance at high aspect ratio. Note, methanol masers become saturated at optical depths higher than about 15. We do not consider saturation effect in our calculations. 

\section{Conclusions}
The calculations of energy level populations of methanol molecule are performed using the results of shock wave modelling by \cite{Nesterenok2018,Nesterenok2019}. We provide the list of methanol transitions that can be Class I methanol masers. Besides the known detected methanol masers, there are many other transitions that may have population inversion and produce masers at specific physical conditions.

\section*{References}
\bibliography{references} 

\end{document}